\documentclass[11pt]{article} 
\usepackage{amsmath,amssymb,graphicx}
\usepackage[sort&compress,numbers]{natbib} 
\usepackage{geometry,amsmath,enumerate}
\newcommand{\CP}{\textsf{CP}}
\newcommand{\tmm}{\mathsf{tm}}
\newcommand{\te}{\mathsf{te}}
\newcommand{\gr}{\mathsf{gr}}
\newcommand{\x}{\textsf{x}}
\newcommand{\ii}{\mathrm{i}}
\renewcommand{\Re}{\mathrm{Re}}
\usepackage{xcolor} \definecolor{darkgreen}{rgb}{0,.5,0}
\usepackage[colorlinks,filecolor=blue,citecolor=darkgreen,unicode]{hyperref}

\title{The low-temperature expansion of the Casimir-Polder free energy of an atom with graphene}

\author{Nail Khusnutdinov\footnote{email: nail.khusnutdinov@gmail.com}\\[1em] 
	\small CMCC, UFABC, 09210-170 Santo Andr\'e, SP, Brazil\\ 
	\small and Regional Scientific and Educational Mathematical Center of Kazan Federal University,	\\ 
	\small Kremlevskaya 18, Kazan, 420008, Russia\\
	\and Natalia Emelianova\footnote{email: natalia.emelianova@ufabc.edu.br}\\[1em]
    \small  CMCC, UFABC, 09210-170 Santo Andr\'e, SP, Brazil}
\date{February 7, 2021}
\begin{document}
	\maketitle

\begin{abstract} 
We consider the low-temperature expansion of the Casimir-Polder free energy for an atom and graphene by using the Poisson representation of the free energy. We extend our previous analysis on the different relations between chemical potential $\mu$ and mass gap parameter $m$. The key role plays the dependence of graphene conductivities on the $\mu$ and $m$. For simplicity, we made the manifest calculations for zero values of the Fermi velocity. For $\mu >m$ the thermal correction $\sim T^2$ and for $\mu < m$ we confirm the recent result of Klimchitskaya and Mostepanenko, that the thermal correction $\sim T^5$. In the case of exact equality $\mu =m$ the correction $\sim T$. This point is unstable and the system falls to the regime with $\mu >m$ or $\mu <m$. The analytical calculations are illustrated by numerical evaluations for the Hydrogen atom/graphene system.
\end{abstract}

\section{Introduction}

The Casimir \cite{Casimir:1948:otabtpcp} and Casimir-Polder \cite{Casimir:1948:TIrLdWf} dispersion forces play an important role in different phenomena \cite{Parsegian:2006:VdWFHBCEP, Bordag:2009:ACE}. The Casimir-Polder force is usually referred to as the van der Waals force on large distances between micro-particles and macro-objects when the retardation of interaction is taken into account.  The Casimir-Polder force essentially depends on the material of the macro-objects, its dimension, shapes, conductivity, and temperature \cite{Parsegian:2006:VdWFHBCEP, Bordag:2009:ACE}, and it is important for the interaction of graphene with micro-particles  \cite{Bondarev:2005:vdWcadcn,Blagov:2007:vdWibmscn,Klimchitskaya:2020:CCPfgsqftdt,Khusnutdinov:2016:CPefasocp,Khusnutdinov:2019:cei2dmmr}. The Casimir-Polder force and torque for anisotropic molecules have been the subject of investigations in the recent years \cite{Babb:2005:lrasica,Marachevsky:2010:CPepCSi,Shajesh:2012:rlrfbaad,Thiyan:2015:acvdWCPeCO2CH4mnstf,Antezza2020cpfatfamctcpateoc}.

The thermal corrections to the Casimir-Polder interaction for micro-particle/graphene were considered in Refs. \cite{Bezerra:2008:Ltaiatqr,Chaichian:2012:TCiodawg,Bordag:2014:lteLf,Khusnutdinov:2018:tccpinp2ddm,Khusnutdinov:2019:lteotcfefaaiwacp,Klimchitskaya:2020:NhtaiwgDmwnegcp,Klimchitskaya:2020:QftdCebtrgst}. In the last few years, much attention is given to the low-temperature expansion of the Casimir-Polder free energy for the atom/graphene system  \cite{Khusnutdinov:2019:lteotcfefaaiwacp,Klimchitskaya:2020:NhtaiwgDmwnegcp,Klimchitskaya:2020:QftdCebtrgst,Klimchitskaya:2020:CCPfgsqftdt}. In the case of an atom/ideal plane, the low-temperature correction to
the Casimir-Polder free energy is proportional to the fourth degree of temperature $\sim T^4$.  As opposed to the ideal case, the conductivity of graphene depends on the chemical potential and temperature, and it has temporal and spatial dispersion   \cite{Bordag:2009:CibapcagdbtDm,Fialkovsky:2011:FCefg,Bordag:2016:ECefdg,Bordag:2017:EECefdg}. The low-temperature expansion depends on the relations between these macro-parameters.

It was shown in Ref. \cite{Khusnutdinov:2019:lteotcfefaaiwacp} that the low-temperature expansion reveals the unusual quadratic $\sim T^2$ behaviour. Next detail considerations \cite{Klimchitskaya:2020:NhtaiwgDmwnegcp,Klimchitskaya:2020:QftdCebtrgst} showed a more rich picture of low-temperature expansion depending on the relation between chemical potential $\mu$ and mass gap parameter $m$ of the Dirac electron. For $\mu >m$ the same quadratic behaviour was confirmed but in the case, $\mu < m$ the $\sim T^5$ dependence was obtained. In the case of the very specific exact relation $\mu =m$ the linear $\sim T$ dependence was observed. To obtain these results the authors of Refs. \cite{Klimchitskaya:2020:NhtaiwgDmwnegcp,Klimchitskaya:2020:QftdCebtrgst} made the sophisticated treatment of the Matsubara series. 

In the present paper, we extend our analysis made in Ref.  \cite{Khusnutdinov:2019:lteotcfefaaiwacp} in the framework of the Poisson representation of the Matsubara series to all relations between $\mu$ and $m$ and confirm results by numerical analysis. In Ref. \cite{Khusnutdinov:2019:lteotcfefaaiwacp} we considered conductivity of graphene with cutting scattering rate as in Refs. \cite{Falkovsky:2007:Sdogc,Gusynin:2007:Mcig} where the Kubo approach was used.  In this case, the conductivity has a constant value at zero frequencies which depends on the scattering rate parameter $\gamma$. It leads to the low-temperature dependence $\sim T^2$ for any relation between chemical potential $\mu$ and mass gap $m$. In the framework of the polarization tensor approach \cite{Fialkovsky:2011:FCefg,Bordag:2016:ECefdg,Bordag:2017:EECefdg}, there is no scattering parameter and the behaviour of the conductivity at zero frequency strongly depends on the relation between $\mu$ and $m$. We show that  for $\mu > m$ the temperature corrections to the free energy $\sim T^2$ and for $\mu < m$ we obtain correction $\sim T^5$ and in the point $\mu=m$ the linear dependence $\sim T$ appears. Therefore, we confirm expansions obtained in Ref.  \cite{Klimchitskaya:2020:NhtaiwgDmwnegcp} in the framework of our approach.    

Throughout the paper the units $\hbar=c=k_B =1$ are used.  
 
\section{The Casimir-Polder free energy}

Taking into account the Poisson summation formula (see details in Ref. \cite{Khusnutdinov:2019:lteotcfefaaiwacp}),  the free energy may be represented in the following form
\begin{equation}
	\frac{\mathcal{F}_{\tmm \left| \te \right.}}{\mathcal{E}_{\CP}^\infty} = - \frac{8}{3}  \sum_{l = 0}^{\infty}{}^{\prime} \int_0^{\infty} z^3 e^{- 2 z} dz 	\int_0^1 dx \cos \left( \frac{zxl}{aT} \right) \frac{\alpha (\lambda)}{\alpha (0)}  \{x^2 - 2 | x^2  \} r^{\tmm \left| \te \right.}, \label{eq:FPoi}
\end{equation}
normalized to the $\mathcal{E}_{\CP}^\infty = - 3 \alpha (0) / 8 \pi a^4$ -- the Casimir-Polder (\CP) energy for an ideal plane/atom at large distance $a$. Here, the prime means factor $1 / 2$ for $l = 0$ and we have to use $\lambda = \frac{zx}{a}, k = \frac{z}{a}  \sqrt{1 - x^2}$ for imaginary frequency $\omega = \ii \lambda$ and wave-vector $k$ in conductivities $\eta^{\tmm} = 2\pi \sigma^\tmm$ and $\eta^{\te} = 2\pi \sigma^\te$. The $\alpha$ is the polarizability of atom or molecule at the imaginary frequency, and the refraction coefficients of TE and TM modes are 
\begin{equation}
	r^{\te}  =  - \frac{1}{1 + \frac{1}{x \eta^\te}}, r^{\tmm}  =  \frac{1}{1 + \frac{x}{\eta^\tmm}} . 
\end{equation}

The form of the zero terms, $l = 0$, in {\eqref{eq:FPoi}} coincides exactly with that obtained for zero temperatures but, in general, with temperature and chemical potential dependence through the conductivities $\eta^{\tmm |\te} = \eta^{\tmm |\te} (\lambda,k,\mu,m,T) $. We extract the zero, $l=0$, term
\begin{equation}\label{eq:Division}
	\mathcal{F}=\mathcal{F}_{\tmm} +\mathcal{F}_{\te} = \mathcal{F}^0 + \Delta \mathcal{F}, 
\end{equation}
and consider the low-temperature expansion for $\Delta \mathcal{F}$ and $\mathcal{F}^0$ separately. We extract the temperature contribution from $\mathcal{F}^0$. Then, 
\begin{equation}
	\mathcal{F}= \mathcal{F}^{T=0} + \Delta \mathcal{F}^0 + \Delta \mathcal{F}, 
\end{equation}    
where $\Delta \mathcal{F}^0 = \mathcal{F}^0  - \mathcal{F}^{T=0}$. Therefore, the total temperature correction, $ \Delta_T \mathcal{F}$, consists of two parts,
\begin{equation}\label{eq:TotalCorrection}
	 \Delta_T \mathcal{F} = \Delta \mathcal{F}^0 + \Delta \mathcal{F}.
\end{equation} 
In general (for the graphene case, for example), the $\mathcal{F}_{T=0}$ depends on the chemical potential.  

The expansion crucially depends on the behaviour of the conductivities at zero frequencies. In Ref. \cite{Khusnutdinov:2019:lteotcfefaaiwacp} we considered in detail the different models of conductivities with a constant value of conductivity at zero frequencies. In particular, we have taken into account the graphen's conductivity with finite scattering factor $\gamma$, which means that the $\eta^{\tmm |\te}|_{\omega \to 0} = \eta^{\tmm |\te}|_{\omega = \gamma}$. As a result, the free energy has the main low-temperature term $\sim T^2$ for any relation between $\mu,m$ and $T$. With zero scattering factor, we have to consider this expansion more carefully.  

In the framework of the polarization tensor approach \cite{Bordag:2009:CibapcagdbtDm}, the conductivities of the TM and TE modes read \cite{Fialkovsky:2011:FCefg,Bordag:2016:ECefdg,Bordag:2017:EECefdg}
\begin{subequations}\label{eq:CondGen}
\begin{equation}
	\eta_i = \eta_i^0 + \Delta \eta_i, \Delta \eta_i = \eta_{\gr} \int_m^{\infty} dyf_i (y) \Xi (y, \mu, T),
\end{equation}
where
\begin{eqnarray}
	\frac{\eta_{\te}^0}{\eta_{\gr}} & = &  \frac{4 m}{\pi \lambda} \left(1 + \frac{k_F^2 - 4m^2}{2mk_F}\arctan \left( \frac{k_F}{2 m} \right) \right), \frac{\eta_{\tmm}^0}{\eta_{\te}^0} = \frac{\lambda^2}{k_F^2} , \nonumber \\
	f_{\te} (y) & = & \frac{8}{\pi \lambda} \Re \frac{(4 m^2 + q^2)  (q^2 k_F^2 		+ 4 m^2 k^2 v_F^2) - q^2 k_F^2 \lambda^2}{r (q^2 k_F^2 + 4 m^2 k^2 v_F^2 + q \lambda r)}, q  =  \lambda - 2 \ii y,\nonumber\\
	f_{\tmm} (y) & = & \frac{8}{\pi} \Re \frac{q (q^2 + k^2 v_F^2 + 4 m^2) - \lambda r}{r (r + q \lambda)}, \Xi = \frac{1}{e^{\frac{y + \mu}{T}} + 1} + 	\frac{1}{e^{\frac{y - \mu}{T}} + 1}, \nonumber\\
	k_F &=& \sqrt{\lambda^2 + v_F^2 k^2}, r = \sqrt{k_F^2  (q^2 + k^2 v_F^2) + 4 m^2 k^2 v_F^2}, 
\end{eqnarray}
\end{subequations}
and $\eta_\gr = \frac{2 \pi \sigma_{\gr}}{c} = \frac{\pi e^2}{2 \hbar c} = 0.0114$ with $ \sigma_\gr = \frac{e^2}{4\hbar}$ being the graphene universal conductivity. 

Let us consider, for simplicity, the conductivity in the zero approximation over the Fermi velocity $v_F = 1/300 \ll 1$. In this approximation, we obtain more simple expressions ($\x = \tmm,\te$)
\begin{equation}
	\frac{\eta_\x^0}{\eta_{\gr}}  =   \frac{4 m}{\pi \lambda} \left(1 + \frac{\lambda^2 - 4m^2}{2m\lambda}\arctan \left( \frac{\lambda}{2 m} \right) \right), f_\x (y)  =  \frac{16 \left(m^2+y^2\right)}{\pi  \lambda  \left(\lambda ^2+4 y^2\right)}, 
\end{equation}
and we observe that the conductivities have no dependence on $k$ which should be the case because the Fermi velocity and wave-vector come in the single combination $kv_F$. 

\subsection{Expansion of the $\Delta \mathcal{F}$}

The sum with $l\geq 1$, the $\Delta \mathcal{F}$, may be represented in the following form \cite{Khusnutdinov:2019:lteotcfefaaiwacp}:
\begin{equation}
	\frac{\Delta \mathcal{F}}{\mathcal{E}_{\CP}^\infty} = \frac{8}{3} \Re \sum_{l = 1}^{\infty} \int_0^{\infty} dz e^{i \Lambda_l z}  \left( Y_{\tmm} + Y_{\te} 	\right), \label{eq:ECaCP}
\end{equation}
where 
\begin{eqnarray}
	Y_{\tmm} (z) & = & \frac{\alpha (\lambda)}{\alpha (0)}  \int_z^{\infty} \frac{e^{- 2 s} s (2 s^2 - z^2)}{s + z / \eta_{\tmm}} ds, \nonumber\\
	Y_{\te} (z) & = & \frac{\alpha (\lambda)}{\alpha (0)}  \int_z^{\infty} \frac{e^{- 2 s} z^3}{z + s / \eta_{\te}} ds, 
\end{eqnarray}
and $\Lambda_l = l/aT, \lambda = z / a$, and $k = \sqrt{s^2 - z^2} / a$. This representation is suitable for $T\to 0$ analysis which means $\Lambda_l \to \infty$. 

In the case $v_F\to 0$, the conductivities do not depend on $k$, and therefore we can calculate integral over $s$:
\begin{eqnarray}
	Y_{\tmm} (z) & = & \frac{\alpha \left( \frac{z}{a} \right) e^{- 2 z}}{\alpha (0) 2 \eta_{\tmm}^3} \left\{ \eta_{\tmm}  \left( \eta_{\tmm}^2 + \left( 	\eta_{\tmm}^2 - 2 \eta_{\tmm} + 2 \right) z^2 + \eta_{\tmm} (2 \eta_{\tmm} - 1) z \right)\right. \nonumber \\
	 &-& \left. 2 \left( \eta_{\tmm}^2 - 2 \right) z^3 e^{2 \left( 1 + 		\eta_{\tmm}^{- 1} \right) z}  \mathrm{Ei} \left[ - 2 z \left( 1 + \eta_{\tmm}^{- 1} \right) \right] \right\}, \nonumber\\
	Y_{\te} (z) & = & - \frac{\alpha \left( \frac{z}{a} \right)}{\alpha (0)} 	\eta_{\te} z^3 e^{2 \eta_{\te} z}  \mathrm{Ei} \left[ - 2 z (1 + \eta_{\te}) 	\right], \label{eq:Yint}
\end{eqnarray}
where $\mathrm{Ei}[x]$ is the exponential logarithm function. The function $\mathrm{Ei}[x]$ has the following representation as a series 
\begin{equation}
\mathrm{Ei} (x) = \gamma_E + \ln (- x) + \sum_{n \geq 1} \frac{x^n}{n \cdot n!},
\end{equation}
that contains polynomials as well as logarithmic contributions.  

Then we make expansion over $z$, 
\begin{equation}
	Y_\x = \sum_{n \geq 0} A_n^\x z^n + \sum_{n \geq 1} B_n^\x z^n \ln z, \label{eq:Yser}
\end{equation}
and use the Lemmas Erd{\'e}lyi\footnote{These lemmas are sometimes called etalon integrals in the asymptotic methods of the stationary phase} (see Ref. \cite[Eqs. 1.13 and 1.35]{Fedoryuk:1977:TSPM} and Ref. \cite{Khusnutdinov:2019:lteotcfefaaiwacp}) to calculate asymptotic $\Lambda_l \to \infty$ for integral over $z$ in Eq. \eqref{eq:ECaCP} for each term of series. The manifest form of the coefficients depends on the specific model of conductivity.  We obtain the series \eqref{eq:Yser} in which we have to make replacements 

\begin{eqnarray*}
	z^n & \rightarrow & n! e^{\ii  \frac{\pi}{2} (n + 1)} \Lambda_l^{- n - 1},\\
	z^n \ln z & \rightarrow & n! e^{\ii  \frac{\pi}{2} (n + 1)}  \left( \frac{\ii \pi}{2} + \psi (n + 1) - \ln \Lambda_l \right) \Lambda_l^{- n - 1},
\end{eqnarray*}
where $\psi(x)$ is the digamma function, and then we take the real part (see Eq. \eqref{eq:ECaCP}) and obtain the following replacements
\begin{eqnarray}
	z^{2 n + 1} & \rightarrow & (- 1)^{n + 1} (2 n + 1) ! \Lambda_l^{- 2 n - 2}, 	\nonumber\\
	z^{2 n} & \rightarrow & 0, \nonumber\\
	z^{2 n + 1} \ln z & \rightarrow & (- 1)^{n + 1} (2 n + 1) ! (\psi (n + 1) - 	\ln \Lambda_l) \Lambda_l^{- 2 n - 2}, \nonumber\\
	z^{2 n} \ln z & \rightarrow & (- 1)^{n + 1} (2 n) ! \frac{\pi}{2} \Lambda_l^{-2 n - 1}.  \label{eq:zz}
\end{eqnarray}

Then we make a summation over $l \geq 1$ and arrive with relation 
\begin{equation}
	\frac{\Delta \mathcal{F}}{\mathcal{E}_{\CP}^\infty} = \frac{8}{3} \left( X_{\tmm} + X_{\te} 	\right), 
\end{equation}
where
\begin{eqnarray*}
	X_\x & = & \sum_{n \geq 0} A_{2 n + 1}^\x  (- 1)^{n + 1} (2 n + 1) ! \zeta_R (2 n + 2) (a T)^{2 n + 2}\\
	& + & \sum_{n \geq 1} B_{2 n}^\x (- 1)^{n + 1} (2 n) ! \frac{\pi}{2} \zeta_R (2 n + 1) (a T)^{2 n + 1}\\
	& + & \sum_{n \geq 0} B_{2 n + 1}^\x (- 1)^{n + 1} (2 n + 1) ! ([\psi (n + 1) + \ln (a T)] \zeta_R (2 n + 2) - \zeta_R' (2 n + 2))  (a T)^{2 n + 2},
\end{eqnarray*}
and $\zeta_R$ is the Riemann zeta function. We observe that the polynomial contributions come from odd $A_{2 n + 1}^\x$ and $B_n^\x$. The logarithmic contributions come from odd $B_{2 n + 1}^i$. The main contribution reads
\begin{eqnarray}
	\frac{\Delta \mathcal{F}_\x}{\mathcal{E}_\CP^\infty} & = & - \frac{4 \pi^2}{9} \left[ A_1^\x - B_1^\x  \left( \gamma_E + 6 \frac{\zeta_R' (2)}{\pi^2 } 	\right) \right] (a T)^2 + \frac{8 \pi \zeta_R (3) }{3} B_2^\x (a T)^3\nonumber \\
	 &+& \frac{8 \pi^4}{45}  \left[ A_3^\x + B_3^\x  \left( 1 - \gamma_E - 90 \frac{\zeta_R' (4)}{\pi^4} \right) \right] (a T)^4  - 32 \pi \zeta_R (5) B_4^\x (a T)^5\nonumber\\
	  &-& \frac{32 \pi^6}{189} \left[ 2 A_5^\x + B_5^\x  \left( 3 - 2 \gamma_E - 1890 \frac{\zeta_R' (6)}{\pi^6} \right) \right] (a T)^6 + \ldots \nonumber\\
	& + & \ln (a T)  \left\{ - \frac{4 \pi^2}{9} B_1^\x (a T)^2 + \frac{8\pi^4}{45}  B_3^\x (a T)^4 - \frac{64 \pi^6}{189} B_5^\x (a T)^6 + \ldots \right\}, 	\label{eq:DF}
\end{eqnarray}
where $\gamma_E$ is the Euler constant. Note, that in general the coefficients $A^\x_n$ and $B^\x_n$ depend on $m,\mu$ and $T$. 

For the constant conductivity case $B_1^\x = B_2^\x = 0$ and $A_1^{\te} = 0, A_1^{\tmm} = - 1 / \left( 2 \eta_{\tmm} \right)$. Therefore, the main contribution comes from TM mode and reads \cite{Khusnutdinov:2019:lteotcfefaaiwacp}
\begin{equation}
\frac{\Delta \mathcal{F}_{}}{\mathcal{E}_{\CP}^\infty} = - \frac{(2\pi a T)^2}{9} A_1^{\tmm} = \frac{2 \pi^2 (a T)^2}{9 	\eta_{\tmm}}.
\end{equation}

\begin{figure}[ht]
	\centering
	\includegraphics[width=7.5 cm]{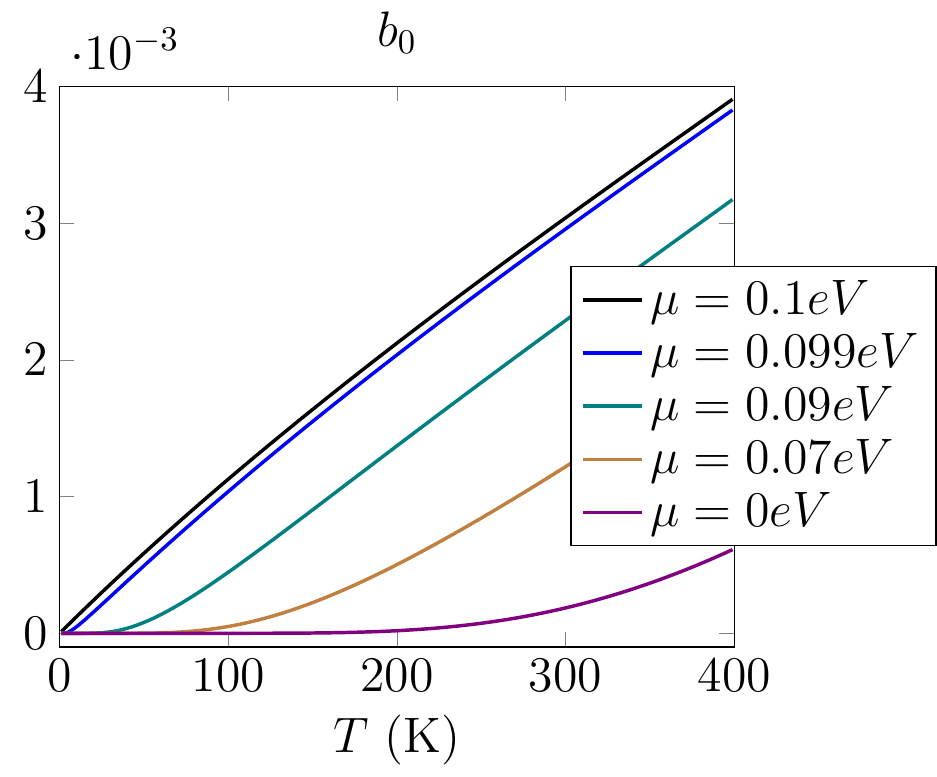}\includegraphics[width=7.5 cm]{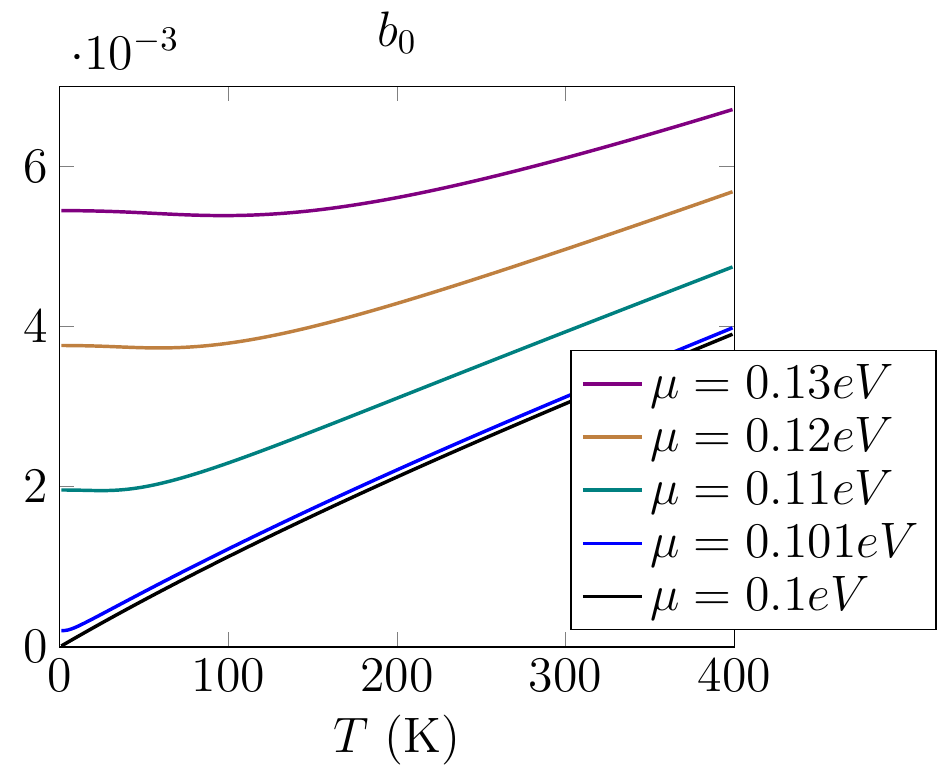}
	\caption{The functions $b_0$ for $a=100$ nm, $m=0.1$eV and different values of chemical potential $\mu$. For $\mu < m$ the function $b_0\sim 0$ for low temperatures; for $\mu = m$ it is linear and for $\mu >m$ it starts from constant values in agreement with Eqs. \eqref{eq:b0Gen}. }\label{fig:-1}
\end{figure}

For graphene case, the conductivities are expanded in the following series over $z$: 
\begin{equation}
\frac{\eta_\x}{\eta_{\gr}} = \frac{1}{z} (b_0 + z^2 b_2 + z^4 b_4 + \ldots),
\end{equation}
where the coefficients $b_n$ are functions of $m,\mu$ and $T$:
\begin{eqnarray}
	b_0 & = & \frac{4 a}{\pi}  \int_m^{\infty} \frac{dy}{y^2}  (m^2 + y^2) \Xi,\nonumber \\
	b_2 & = & \frac{4}{3 \pi m a} - \frac{1}{\pi a}  \int_m^{\infty} \frac{dy}{y^4} (m^2 + y^2) \Xi,\nonumber \\
	b_4 & = & - \frac{2}{15 \pi m^3 a^3} + \frac{1}{4 \pi a^3}  \int_m^{\infty} \frac{dy}{y^6}  (m^2 + y^2) \Xi . 
\end{eqnarray}
The zero term $b_0$ crucially depends on $m$ and $\mu$ for low temperatures:\\
\begin{subequations}\label{eq:b0Gen}
I. $\mu > m, T \ll \mu - m$
		\begin{equation}\label{eq:SmalT1}
			b_0 = \int_m^{\mu} g (y) dy + \frac{\pi^2}{6} T^2 g' (\mu) + \frac{7 \pi^4}{360} T^4 g''' (\mu) + O (e^{- \frac{\mu - m}{T}}) .
		\end{equation}
II. $\mu = m, T \ll m$
		\begin{equation}\label{eq:SmalT2}
			b_0 = T \ln 2 g (m) + \frac{\pi^2}{12} T^2 g' (m) + \frac{3}{4} \zeta_R (3) T^3 g'' (m) + \frac{7 \pi^4}{720} T^4 g''' (m) + O (e^{-				\frac{m}{T}}),
		\end{equation}
III. $\mu < m, T \ll m - \mu$
		\begin{equation}\label{eq:SmalT3}
			b_0 = O (e^{- \frac{m - \mu}{T}}) .
		\end{equation}
\end{subequations}
Here, $g(y) = \frac{4 a}{\pi}  \frac{m^2 + y^2}{y^2}$. These asymptotic are illustrated by numerical evaluation in Figure\,\ref{fig:-1}. 

In the first case \eqref{eq:SmalT1} with $\mu > m$, the expansion $b_0$ over $T$ starts from the constant term (see Figure\,\ref{fig:-1}, right panel) and we obtain the following non-zero coefficients
\begin{eqnarray}
	A_3^{\tmm} & = & \frac{1}{3},\ A_3^{\te}  =  - 1,\nonumber \\
	A_5^{\tmm} & = & \frac{\alpha'' (0)}{6 a^2 \alpha (0)} + \frac{2}{3\eta_\gr b_0} - 	\frac{3}{\eta_\gr^2 b_0^2} - \frac{2}{15},\ A_5^{\te}  =  - \frac{\alpha'' (0)}{2 a^2 \alpha (0)} - \frac{2}{3 \eta_\gr b_0} - \frac{1}{3 \eta_\gr^2 b_0^2} - \frac{2}{3}, \nonumber \\
	B_4^{\tmm} & = & - \frac{1}{\eta_\gr b_0},\ B_6^{\tmm}  =  - \frac{\alpha'' (0)}{2 a^2 \alpha (0) \eta_\gr b_0} + \frac{b_2 - 2}{\eta_\gr^2 b_0^2} + \frac{2}{\eta_\gr^3 b_0^3} .
\end{eqnarray}
The main contribution to the $b_0$ comes from the first term of expansion  in Eq. \eqref{eq:SmalT1}:
\begin{equation}
	b_0 = \frac{4a}{\pi} \frac{\mu^2 -m^2}{\mu}.
\end{equation} 

In the last case \eqref{eq:SmalT3} with $\mu < m$, the $b_0$ is exponentially small (see Figure\,\ref{fig:-1}, left panel) and we set it zero and the conductivities are expanded as the following
\begin{equation}
	\frac{\eta_\x}{\eta_{\gr}} = z 	( b_2 + z^2 b_4 + \ldots), \label{eq:eta}
\end{equation} 
and the firsts non-zero coefficients read
\begin{equation}
	A_5^{\tmm}  =  \frac{1}{15} \eta_\gr b_2  (\eta_\gr b_2 + 2),\ A_5^{\te}  =  2  \eta_\gr b_2,\ B_4^{\te}  =  - \eta_\gr b_2.
\end{equation}

For the specific case \eqref{eq:SmalT2}, when $\mu = m$, we consider the main contribution for $T\to 0$ and can set $b_0=0$ as in the case \eqref{eq:SmalT3}. Therefore, for all cases, we have the following expansion up to $T^5$:
\begin{eqnarray}
	\frac{\Delta \mathcal{F}}{\mathcal{E}_{\CP}^\infty} & = & - \frac{16 \pi^4}{135} (a T)^4 + \frac{8 \pi^2 \zeta_R (3) \mu}{a\eta_\gr (\mu^2 - m^2)} (a T)^5 + \ldots,\  (\mu > m, T \ll \mu - m),\nonumber \\
	\frac{\Delta \mathcal{F}}{\mathcal{E}_{\CP}^\infty} & = & \frac{128 \eta_\gr \zeta_R  (5)}{3 a m} (a T)^5 + \ldots, (\mu = m, T \ll m), \nonumber \\
	\frac{\Delta \mathcal{F}}{\mathcal{E}_{\CP}^\infty} & = & \frac{128 \eta_\gr \zeta_R 	(5)}{3 a m} (a T)^5 + \ldots, (\mu < m, T \ll m - \mu) .\label{eq:DeltaF}
\end{eqnarray}

\subsection{Expansion of the $\mathcal{F}^0$}

The zero term reads
\begin{equation}
	\frac{\mathcal{F}_{\tmm |\te}^0}{\mathcal{E}_{\CP}^\infty}  =  \frac{4}{3} \int_0^{\infty} z^3 e^{- 2 z} dz \int_0^1 dx \frac{\alpha (\frac{zx}{a})}{\alpha (0)}  \left\{\left.\frac{2 - x^2}{1 + \frac{x}{\eta^{\tmm}}} \right|  \frac{x^2}{1 + \frac{1}{x \eta^{\te}}}\right\}.
\end{equation}
The conductivities \eqref{eq:CondGen} depend on the temperature and $\Delta\eta_\x$  have the following low-temperature expansions \cite{Khusnutdinov:2019:lteotcfefaaiwacp}\\
\begin{subequations} \label{eq:etagen}
	I. $\mu > m, T \ll \mu - m$
	\begin{equation}
		\frac{\Delta\eta_\x}{\eta_{\gr}} = \int_m^{\mu} f_\x (y) dy + \frac{\pi^2}{6} T^2 f_\x' (\mu) + \frac{7 \pi^4}{360} T^4 f_\x''' (\mu) + O (e^{- \frac{\mu - m}{T}}) .
	\end{equation}
	II. $\mu = m, T \ll m$
	\begin{equation}
		\frac{\Delta\eta_\x}{\eta_{\gr}} = T \ln 2 f_\x (m) + \frac{\pi^2}{12} T^2 f_\x' (m) + \frac{3}{4} \zeta_R (3) T^3 f_\x'' (m) + \frac{7 \pi^4}{720} T^4 f_\x''' (m) + O (e^{-\frac{m}{T}}),
	\end{equation}
	III. $\mu < m, T \ll m - \mu$
	\begin{equation}
		\frac{\Delta\eta_\x}{\eta_{\gr}}  = O (e^{- \frac{m - \mu}{T}}) .
	\end{equation}
\end{subequations}

Taking these expansions into account we obtain the following low-temperature corrections to zero term 
\begin{eqnarray}
\frac{\Delta\mathcal{F}_\x^0}{\mathcal{E}_{\CP}^\infty} & = & (a T)^2 G_\x, (\mu > m, T \ll \mu - m),\nonumber \\
\frac{\Delta \mathcal{F}_\x^0}{\mathcal{E}_{\CP}^\infty} & = & (a T) H_\x, (\mu = m, T \ll m),\nonumber \\
\frac{\Delta \mathcal{F}_\x^0}{\mathcal{E}_{\CP}^\infty} & = & 0, (\mu < m, T \ll \mu - m),
\end{eqnarray}
where
\begin{eqnarray}
	G_{\tmm |\te} & = & \frac{2 \pi^2}{9 a^2} \int_0^{\infty} z^3 e^{- 2 z} dz \int_0^1 dx \frac{\alpha (\frac{z x}{a})}{\alpha (0)}  \left\{\left.\frac{(2 - x^2) x f'_\tmm (\mu)}{(x + \eta_{\tmm}^0 + \Delta_0 \eta_{\tmm})^2} \right| \frac{x^3 f'_\te (\mu)}{(1 + x \left( \eta_{\te}^0 + \Delta_0 \eta_{\te} \right))^2}\right\} ,\nonumber \\
	H_{\tmm |\te} & = & \frac{4 \ln 2}{3 a} \int_0^{\infty} z^3 e^{- 2 z} dz \int_0^1 dx \frac{\alpha (\frac{z x}{a})}{\alpha (0)}  \left\{\left. \frac{(2 - x^2) xf_\tmm (m)}{(x +\eta_{\tmm}^0)^2} \right| \frac{x^3 f_\te (m)}{(1 + x	\eta_{\te}^0)^2}\right\}.
\end{eqnarray}
The functions $f_\x$ are given by Eq. \eqref{eq:CondGen} and
\begin{equation}
	\Delta_0 \eta_\x = \eta_{\gr} \int_m^\mu f_\x(y)dy.
\end{equation} 

Therefore, taking into account \eqref{eq:DeltaF} we obtain the low-temperature expansion of the free energy
\begin{subequations}\label{eq:DeltaTotal}
	\begin{eqnarray}
	\frac{\Delta_T\mathcal{F}}{\mathcal{E}_{\CP}^\infty} & = & (a T)^2 (G_\tmm + G_\te), (\mu > m, T \ll \mu - m),\label{eq:DeltaTotal-1} \\
	\frac{\Delta_T \mathcal{F}}{\mathcal{E}_{\CP}^\infty} & = & (a T) (H_\tmm + H_\te), (\mu = m, T \ll m),\label{eq:DeltaTotal-2}\\
	\frac{\Delta_T\mathcal{F}}{\mathcal{E}_{\CP}^\infty} & = & \frac{128 \eta_\gr\zeta_R 	(5)}{3 a m} (a T)^5 + \ldots. (\mu < m, T \ll m - \mu). \label{eq:DeltaTotal-3}
\end{eqnarray}
\end{subequations}

\section{Numerical evaluations}

We evaluated numerically the total temperature correction \eqref{eq:TotalCorrection} by using the expression for the free energy in the form of Matsubara sum. We considered the Hydrogen atom at distance $a=100$ nm from the graphene sheet. The polarizability of the Hydrogen atom in the single oscillator approximation may be found in Ref. \cite{Khusnutdinov:2016:CPefasocp}, for example. The graphene conductivities are given by Eqs. \eqref{eq:CondGen}. We used the Fermi velocity $v_F=1/300$ and the mass gap $m=0.1$ eV. To visualize the dynamic of $\mu$ dependence we made calculations for the value of $\mu$ close to $m=0.1$ eV. The free energy is normalized to the $\mathcal{E}_\CP^\infty = - 3 \alpha (0) / 8 \pi a^4$ -- the \CP energy for an ideal plane/atom at large distance $a$.

\begin{figure}[ht]
	\centering
	\includegraphics[width=7.5 cm]{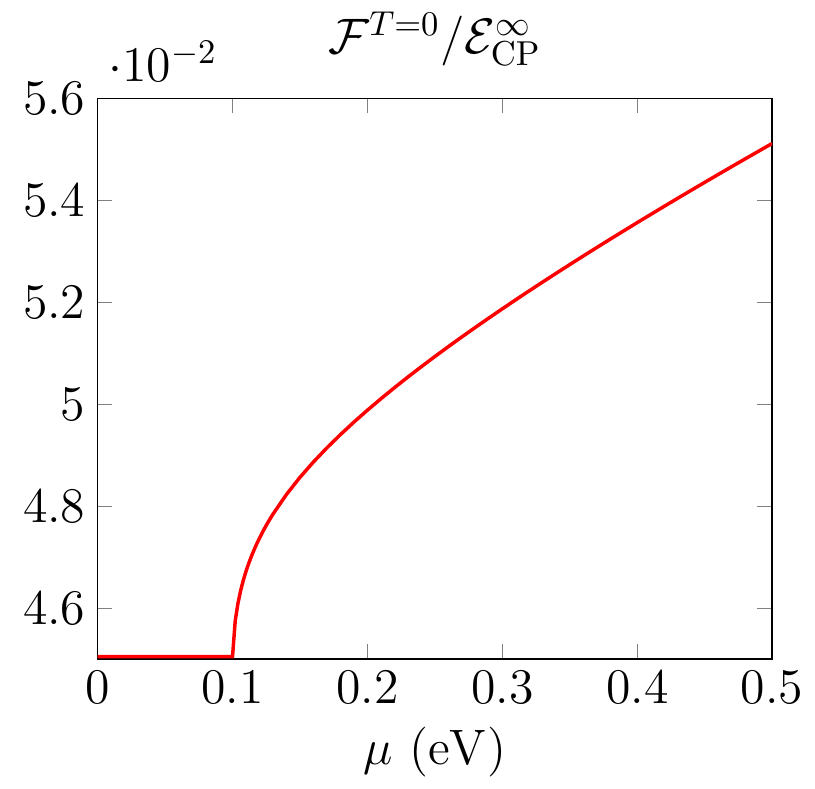}
	\caption{The zero temperature free energy, $\mathcal{F}^{T=0}$, as a function of chemical potential $\mu$.}\label{fig:FT0}
\end{figure}

The zero-temperature free energy, $\mathcal{F}^{T=0}$, depends on the chemical potential $\mu$ and this dependence is shown in Figure\,\ref{fig:FT0}. For $\mu \leq m$ it has the constant value, and it grows up starting with mass gap $\mu >m$.   

\begin{figure}[ht]
	\centering
	\includegraphics[width=7.5 cm]{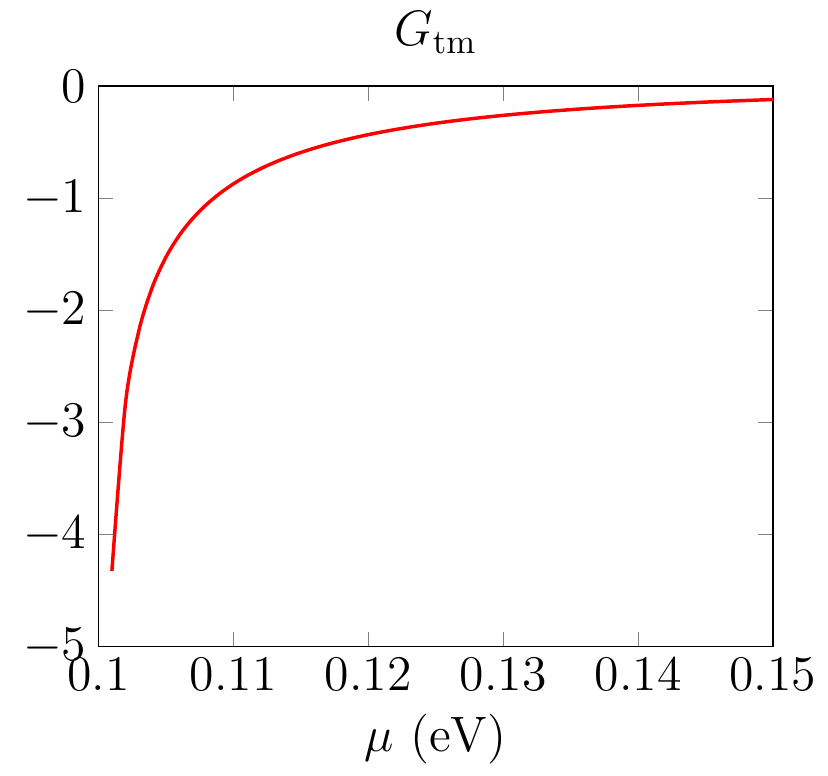}\includegraphics[width=7.5 cm]{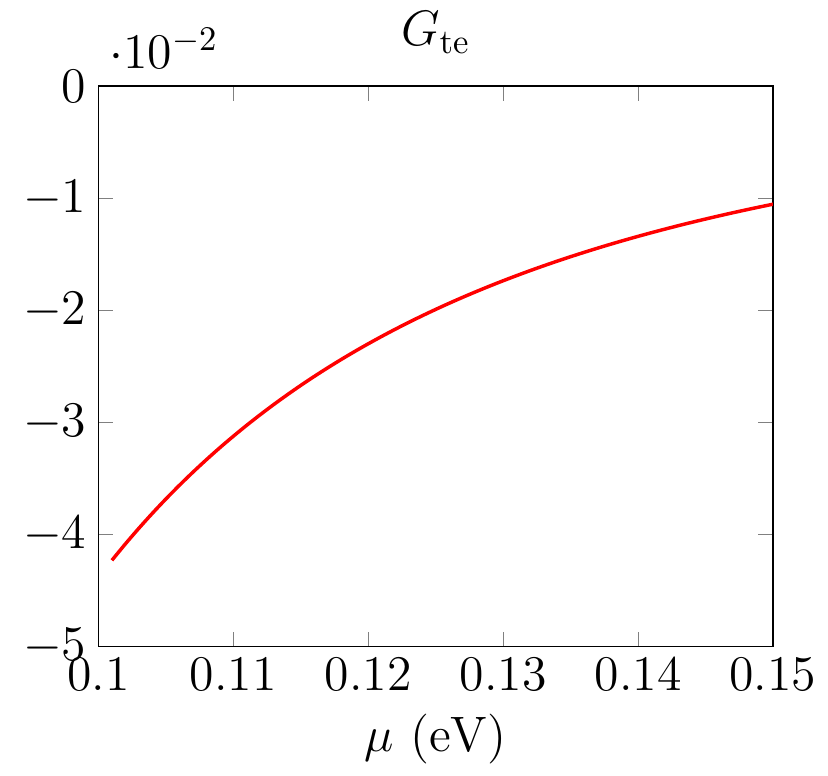}
	\caption{The functions $G_\x$ \eqref{eq:DeltaTotal-1} for different values of the chemical potential $\mu > m$.}\label{fig:0}
\end{figure}

We proceed now to the consideration of the temperature correction, $\Delta_T \mathcal{F}$ to the free energy. First of all let us consider the functions $G_\x$ which define low temperature expansion for $\mu>m$ case \eqref{eq:DeltaTotal-1}. They are plotted in Figure\,\ref{fig:0}. We observe that they are negative and contribution from the TE mode is $100$ times smaller. 

The numerical evaluation of $H_\x(a)$ for $a=100$ nm gives the following values $H_\tmm = 3\cdot 10^6$ and  $H_\te = 0.18$. Again the main contribution comes from TM mode. The value of $H_\tmm$ strongly depends on the value of the Fermi velocity. For $v_F\to 0$ the $H_\tmm \to \infty$. 

\begin{figure}[ht]
	\centering
	\includegraphics[width=6.5 cm]{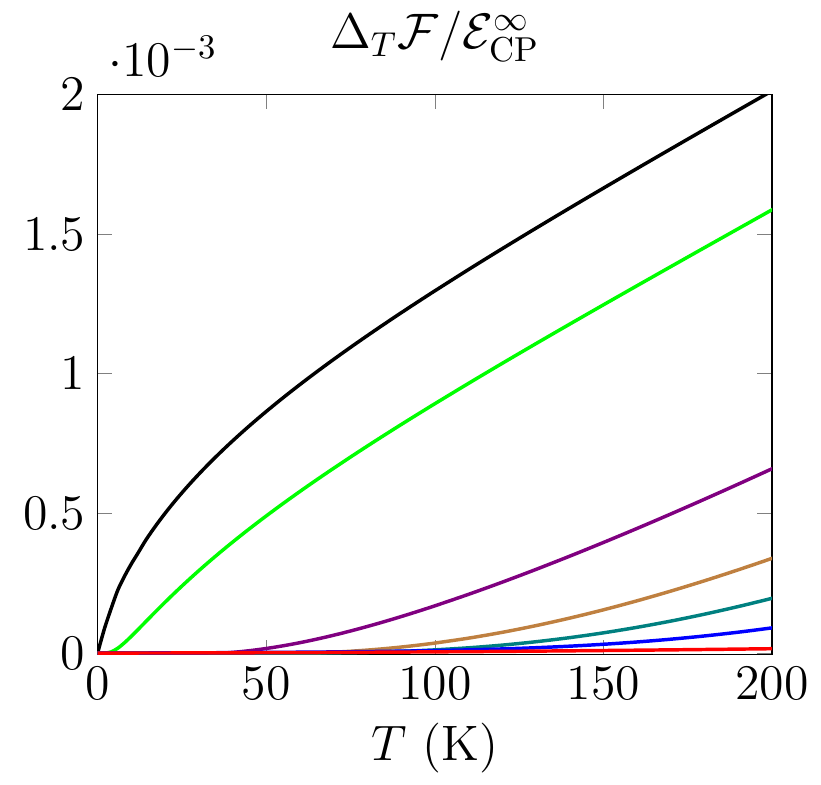}\includegraphics[width=7.5 cm]{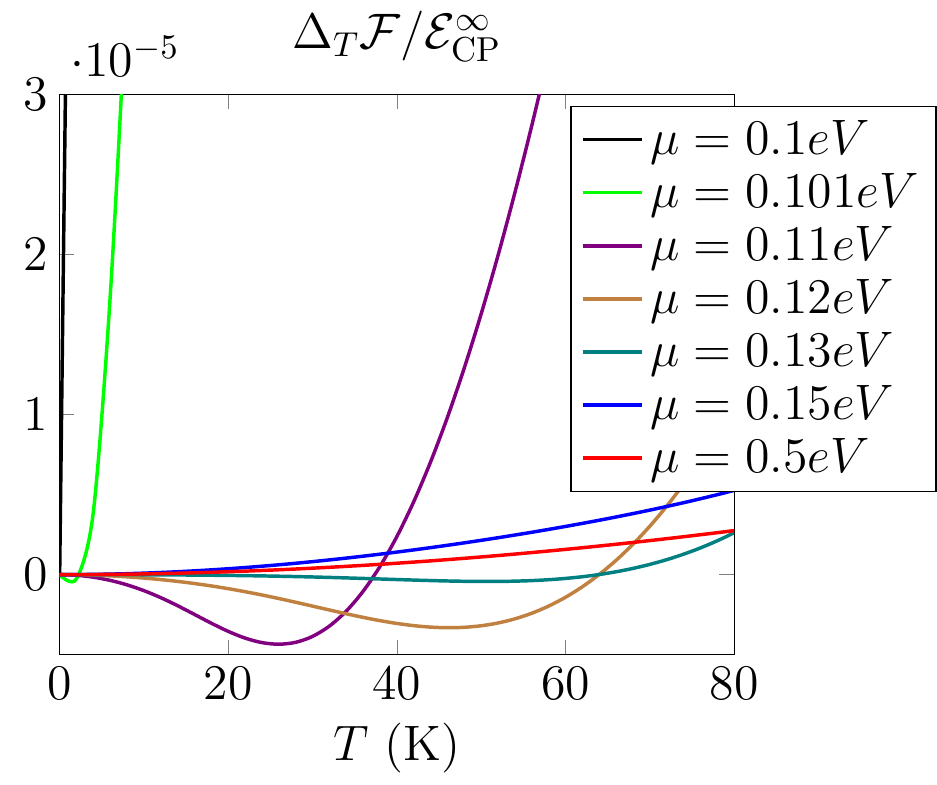}
	\caption{The temperature contribution to the free energy in the different intervals of temperatures and $\mu \geq m$. }\label{fig:1}
\end{figure}

The temperature contribution for $\mu \geq m$ is shown in Figure\,\ref{fig:1}. We observe that for low temperatures the free energy has the form of parabola, $\sim G_\tmm (aT)^2$, in agreement with \eqref{eq:DeltaTotal} with negative parameter $G_\tmm$ (see Figure\,\ref{fig:0}).

The closer $\mu >m$ to $m$, the smaller domain of temperature $T$ where this approximation valid, and the greater value of the parameter of parabola $G_\tmm$, in agreement with Figure\,\ref{fig:0}. If $\mu=m$ this domain becomes zero and the free energy changes drastically its form. If $\mu\to 0$, the part of the curve which is out of this domain (the vertical part of the green curve, for example) goes to free energy for this very special position with $\mu=m$ (black curve). Therefore, for any infinitely small difference $\mu -m \not = 0$ the derivative of free energy with respect temperature $T$ is zero for $T=0$ and the Nernst theorem is valid. The experimental realization of the exact equality $\mu =m$ can not be realized,  and we conclude that the Nernst theorem is valid for this system.   

\begin{figure}[ht]
	\centering
	\includegraphics[width=7.0 cm]{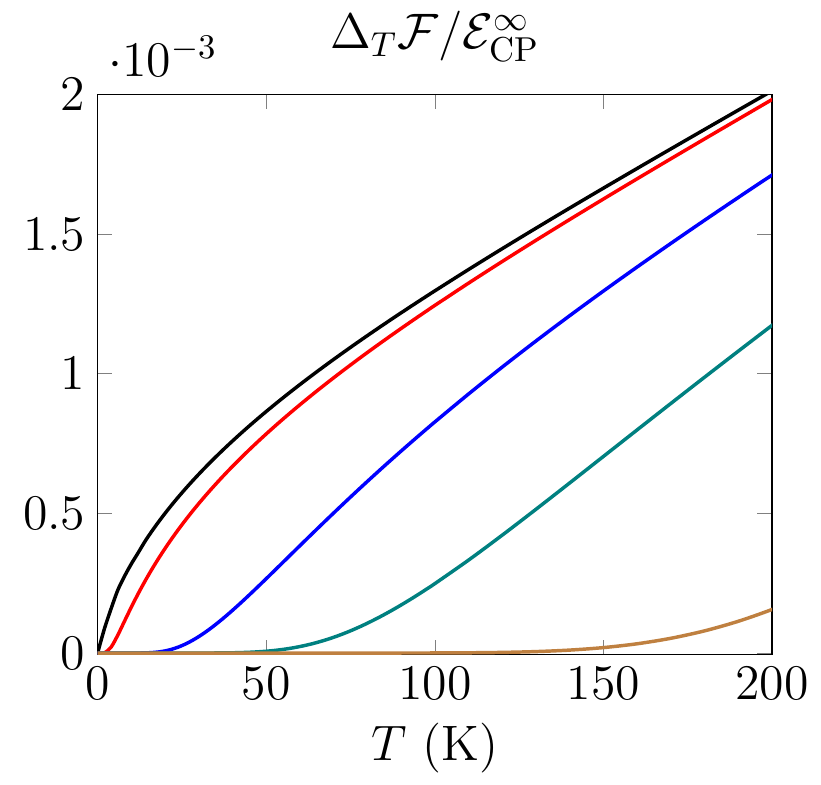}\includegraphics[width=8 cm]{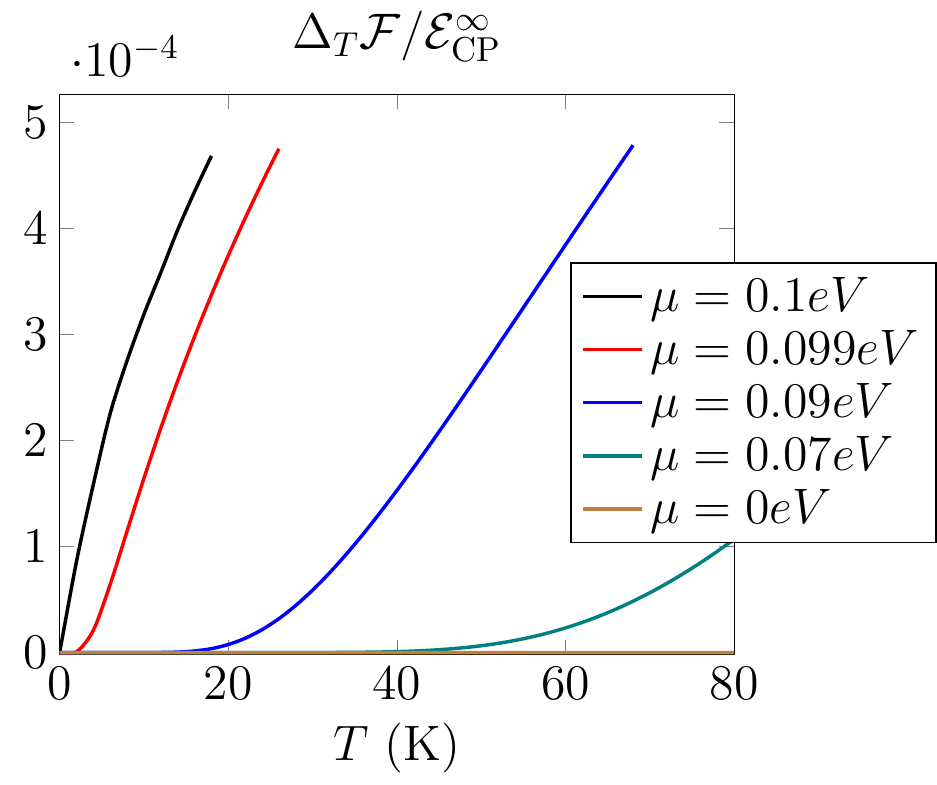}
	\caption{The temperature contribution to the free energy in the different intervals of temperatures and $\mu \leq m$.}\label{fig:2}
\end{figure}

The temperature contribution for $\mu \leq m$ is shown in Figure\,\ref{fig:2}. We observe the completely different dependence of the energy on the $\mu$.   

For zero chemical potential (brown curve), the temperature correction is, in fact, zero for the large domain of temperatures.   The closer $\mu$ to $m$, the smaller domain in which temperature correction is zero. According to \eqref{eq:DeltaTotal-2}
\begin{equation}
		\frac{\Delta_T\mathcal{F}}{\mathcal{E}_{\CP}^\infty}  = \frac{128 \eta_\gr\zeta_R 	(5)}{3 a m} (a T)^5  = 10^{-24} T^5(K),
\end{equation}
in this domain. 

\begin{figure}[ht]
	\centering
	\includegraphics[width=7.0 cm]{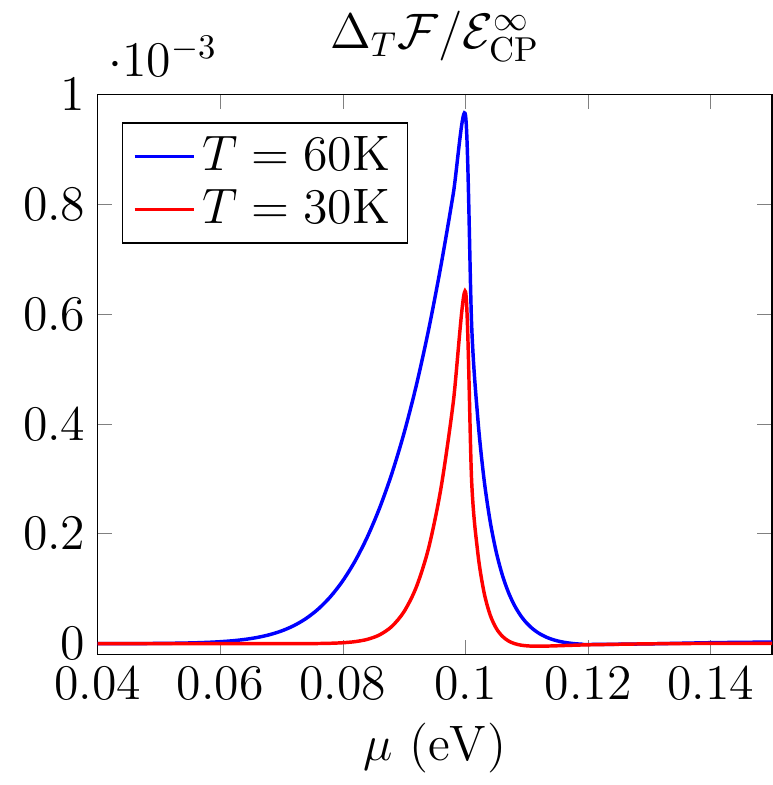}
	\caption{The temperature contribution to the free energy as a function of the chemical potential for distance between Hydrogen atom and  graphene $a=100$nm. }\label{fig:3} 
\end{figure}

The Figure\,\ref{fig:3} shows the temperature contribution to the free energy as a function of chemical potential. The function has a very sharp form with the maximum for $\mu = m$ with a different slope at the left and the right of this point.  

The point  $\mu =m$ looks like phase transition point between different regimes from $\Delta_T \mathcal{F} \sim T^4$ to $\Delta_T \mathcal{F} \sim T^2$. In fact, it is an unstable point -- infinitely small deviation $\mu$ from $m$ changes regime.     

From the Figures\,\ref{fig:1},\ref{fig:2} and relations \eqref{eq:DeltaTotal} we observe the different signs of the entropy, $S$, for $\mu < m$ and $\mu >m$. The entropy is the negative derivative of the free energy with respect of temperature. Therefore, $S^{\mu < m} <0$ (see Figure\,\ref{fig:2}) and $S^{\mu > m} > 0$ (see Figure\,\ref{fig:1}).   In both case the Nernst theorem is valid, $S_{T\to 0} \to 0$. The negative entropy of the dispersion forces has already been observed in Refs. \cite{Bezerra:2008:Ltaiatqr,Khusnutdinov:2012:tcpiawsps} for plain and spherical configurations in the framework of plasma model and also was discussed recently in Ref. \cite{Li:2021:notcseisg}. 

\section{Conclusions}

We considered the low-temperature correction to the Casimir-Polder free energy for atom/graphene system by using the Poisson representation of the free energy, which is more suitable for low-temperature analysis. The analysis is naturally broken into three different regions: i) $\mu >m$, ii) $\mu =m$ and iii) $\mu <m$ for chemical potential. This division is the consequence of the same regions for the conductivity of graphene (see Eq. \eqref{eq:etagen}). The conductivities have completely different expansion in these regions. It starts from the constant in the first region, linear on the temperature in the second one, and exponential small in the third region.  

The free energy may be divided into the two parts \eqref{eq:Division}. The first one, $\mathcal{F}^0$, has the form of the free energy at zero temperature but with $\mu,m$ and $T$ dependence via the conductivity dependence on these parameters. The main contribution in the low-temperature expansion in the first (i) and second (ii) regimes comes from this first term, and it is quadratic and linear on the temperature correspondingly (see Eq. \eqref{eq:DeltaTotal}). In the third (iii) regime, the main contribution $\sim T^5$ comes from the rest part $\Delta \mathcal{F}$.

\section*{Acknowledgments}
We are grateful to Dmitri Vassilevich, Galina Klimchitskaya, and Vladimir Mostepanenko for fruitful discussions.  The NK was supported in part by the grants 2019/10719-9, 2016/03319-6 of São Paulo Research Foundation (FAPESP) and by the Russian Foundation for Basic Research Grant No. 19-02-00496-a.

%

\end{document}